\documentclass[12pt]{article}

\begin{document}

\title{\bf\Large{Thermodynamics of Classical Systems
on Noncommutative Phase Space}}

\author{\\Mojtaba Najafizadeh\\
\it\small{Department of Physics, Faculty of Sciences,
 Tarbiat Modares University,}\\
\it\small{P.O.Box: 14115-111, Tehran, Iran.}\\
\it\small{mnajafizadeh@gmail.com}\\\\
and\\\\
Mehdi Saadat\\
\it\small{Department of Physics, Faculty of Sciences, Shahid Rajaee Teacher}\\
\it\small{ Training University,}\\
\it\small{P.O.Box: 16785-163, Tehran, Iran.} \\
\it\small{saadat@srttu.edu}}
\maketitle

\begin{abstract}
We study the formulation of statistical mechanics on 
noncommutative classical phase space, and construct the
corresponding canonical ensemble theory. For illustration, some basic and important
examples are considered in the framework of noncommutative statistical mechanics:
such as the ideal gas, the extreme relativistic gas and the 3-dimensional
harmonic oscillator.
{\it{keywords}}: Noncommutative Phase-Space; 
Statistical Mechanics; Partition Function.

\end{abstract}

\section{\large{Introduction}}

In recent years, there has been an increasing interest
in the study of physics on noncommutative (NC) spaces, because
the effects of the space noncommutativity may
become significant under extreme conditions, such as at 
energies above the TeV scale, or even at the string scale. 
There are many papers devoted to the study of
various aspects of quantum field theory and quantum
mechanics on NC spaces, where space coordinates are noncommuting 
with each other, but the momenta are commuting,
or on NC phase space, where both ``space-space'' and
``momentum-momentum'' commutation relations can be 
nonvanishing. For references,
see [1]-[10]. The Bose-Einstein statistics of noncommutative
quantum mechanics requires both space-space and
momentum-momentum noncommutativity [11, 12]. On the NC phase space, 
the NC algebra can be written as
\begin{equation}
    [\hat{x_i},\hat{x}_j]=i\hbar\theta_{ij},\quad
    [\hat{p_i},\hat{p}_j]=i\hbar\bar{\theta}_{ij},\quad
    [\hat{x_i},\hat{p}_j]=i\hbar(\delta_{ij}-{1\over 4}\theta_{ik}\bar{\theta}_{kj}) \, ,
\end{equation}
where $\theta_{ij}$  is related to the noncommutativity of the
space coordinates, while $\bar{\theta}_{ij}$ 
 reflects the noncommutativity of the
momenta, and both of them are antisymmetric matrices with real
constant elements. From the relations above, one can obtain a generalized
Bopp's shift as
\begin{eqnarray}
\hat{x}_i & = & x_i-{1\over2}\theta_{ij}p_j,\quad      \\
\hat{p}_i & = & p_i+{1\over2}\bar\theta_{ij}x_j,\quad
\end{eqnarray}
where $x_i$ and $p_i$ are the coordinate and momentum
operators on the usual (commutative) phase 
space, and $i,j=1,2,3$. After applying this shift, the effect
caused by phase space noncommutativity can be calculated
in the usual phase space [13]. In NC quantum mechanics and NC quantum field theory,
the star product between two fields on NC phase space
can be replaced by the generalized Bopp's shift (2) for
coordinates and (3) for momenta. The star product on the NC
 phase space can be defined as [13] 
\begin{eqnarray}
(f \ast g)(x,p) & = & f(x,p)\; e^{  {i\hbar\over2} 
{\overleftarrow{{\partial}^{x}_{i}}} 
{\theta_{ij}} 
\overrightarrow{{\partial}^{x}_{j}}+{i\hbar\over2} \overleftarrow{{\partial}^{p}_{i}}{\bar\theta_{ij}} 
\overrightarrow{{\partial}^{p}_{j}}             }  \; g(x,p) \nonumber \\
& = & f(x,p) g(x,p) + {i\hbar\over2}{\theta_{ij}} {\partial}^{x}_{i}
 f(x,p) {\partial}^{x}_{j} g(x,p) \nonumber \\
& + & {i\hbar\over2}{\bar\theta_{ij}} {\partial}^{p}_{i}
 f(x,p) {\partial}^{p}_{j} g(x,p) + O(\theta ^2)+ O(\bar\theta ^2).
\end{eqnarray}
In this work, in Section 2,
we present the uncertainty relations in the NC phase space
and work out the NC deformation of Planck's constant, 
which leads to a dimensionless classical partition 
function. In Section 3,
assuming the existence of a symplectic structure
consistent with the commutation rules (1), the
 corresponding classical partition function is derived. In Sections 4 to 
6, three concrete examples are presented: the classical ideal gas, the extreme
relativistic gas and the classical harmonic oscillator in a
3-dimensional NC phase space, and the new features that arise are discussed.

\section{\large{The Uncertainty
 Relation on NC Phase Space}}
In the 3-dimensional classical partition function
for a single particle, we put $1\over h^3 $ as the quantity
which makes the volume of phase space dimensionless.
In this section, we derive the analogous factor, that makes
 the volume of 
NC phase space dimensionless.  
From the relation (1), one can write
\begin{equation}
     [\hat{x_i},\hat{p}_j]=i\tilde{\hbar}_{ij},
\end{equation}
where  $\tilde{\hbar}_{ij}$  is a tensor playing the role of the deformation of Planck's
 constant on a NC phase space. By setting  $\theta_3=\theta$ and $\bar{\theta}_3=\bar{\theta}$,
and the rest of the $\theta$  and  $\bar{\theta}$ components to zero
(which can be done by a rotation or a redefinition of coordinates),
Eq. (5) can be rewritten as
\begin{equation}
[\hat{x_i},\hat{p}_j]=i\hbar{\left(\begin{array}{ccc}
1+{\theta\bar{\theta}\over 16} & 0 & ~ 0 \\
0 & 1+{\theta\bar{\theta}\over 16} & ~ 0 \\
0 & 0 & ~ 1 
\end{array}\right)},~~~~~~~~~~~~i,j=1,2,3
\end{equation}
where $\theta_{ij}={1\over2} {\epsilon_{ijk}\theta_k}$ and $\bar{\theta}_{ij}={1\over2}{\epsilon_{ijk}\bar{\theta}_{k}}$.
The following equations
\begin{eqnarray}
   {[\hat{x}_1,\hat{p}_1]} & = & i\tilde{\hbar}_{11} =  i\hbar{ \bigg(1+{\theta\bar{\theta}\over 16} \bigg) },    \\
   {[\hat{x}_2,\hat{p}_2]} & = & i\tilde{\hbar}_{22} =  i\hbar { \bigg(1+{\theta\bar{\theta}\over 16}\bigg) },    \\
   {[\hat{x}_3,\hat{p}_3]} & = & i\tilde{\hbar}_{33} =  i\hbar~,        
\end{eqnarray}
lead to the uncertainty relations on NC phase space
\newpage
\begin{eqnarray}
  { \Delta{\hat{x}_1}\Delta{\hat{p}_1} } & \sim & \tilde{h}_{11} = h\bigg(1+{\theta\bar{\theta}\over 16}\bigg),  \\
  { \Delta{\hat{x}_2}\Delta{\hat{p}_2} } & \sim & \tilde{h}_{22} = h\bigg(1+{\theta\bar{\theta}\over16}\bigg),   \\
  { \Delta{\hat{x}_3}\Delta{\hat{p}_3} } & \sim & \tilde{h}_{33} = h.
\end{eqnarray}
From these relations, it is obvious that the following
 factor makes the volume of
NC phase space dimensionless
\begin{equation}
 { 1\over{\tilde{h}_{11}\tilde{h}_{22}
\tilde{h}_{33}} }={1\over{{\tilde{h}}^3}}=
{1\over  {h^3(1+ {\theta\bar{\theta}\over8 })  }}.
\end{equation}
In this work, we considered an expansion of the denominator up to the second order
 in the NC parameters. It should be mentioned that the deformation of 
the Planck constant, on a 2-dimensional
NC phase space, has the form [13]
\begin{equation}
\tilde{h}=h(1+ {\theta\bar{\theta}\over4 }  ),
\end{equation}
where $\theta_{ij}= {\epsilon_{ij}\theta}$ and $\bar{\theta}_{ij}={\epsilon_{ij}\bar{\theta}}$. Thus, 
the appropriate factor, that makes the volume 
of the NC phase space dimensionless, can 
be written as 
\begin{equation}
{1\over{{\tilde{h}}^2}}=
{1\over  {h^2(1+ {\theta\bar{\theta}\over2 })  }}.
\end{equation}

\section{\large{Classical Partition Function on NC Phase Space}}
The purpose of this paper is precisely to study noncommutative
 classical systems. The passage between NC classical mechanics
and NC quantum mechanics is assumed to be realized via the
 following generalized Dirac quantization condition: 
\begin{equation}
\{f,g\} \longrightarrow {1\over {i\hbar}}[O_f,O_g]
\end{equation}
where we denote the operator
 associated with a classical observable $f$ as $O_f$.
This quantization generalizes the relations
 (1) in the following way:
\begin{equation}
    \{\hat{q_i},\hat{q}_j\}=\theta_{ij}~,\quad
    \{\hat{p_i},\hat{p}_j\}=\bar{\theta}_{ij}~,\quad
    \{\hat{q_i},\hat{p}_j\}=\delta_{ij}-{1\over4}\theta_{ik}
\bar{\theta}_{kj}~, 
\end{equation}
where $\hat{q_i}$ and
 $\hat{p_i}$ are the coordinate and momentum 
classical observables in NC phase space. Moreover,
 the dimensions of $\theta_{ij}$ and $\bar{\theta}_{ij}$ are 
$(\it{length})^2/\hbar$ and $(\it{momentum})^2/\hbar$
respectively, with $i,j = 1,2,3$. 
From the relations above, one can derive an expression for  
NC classical observables as
\begin{eqnarray}
\hat{q}_i & = & q_i-{1\over2}\theta_{ij}p_j,\quad      \\
\hat{p}_i & = & p_i+{1\over2}\bar\theta_{ij}q_j,\quad
\end{eqnarray}
where $q_i=(x,y,z)$ and $p_i=(p_x,p_y,p_z)$ are
 the classical observables in the usual (commutative) phase 
space. As one can see, in the classical limit,
the symplectic structure (17) will not depend on $\hbar$, 
as expected.\\
To obtain the classical partition function, in the canonical ensemble
 in NC phase space,
 it is possible to consider the following 
formula
\begin{equation}
Q_1^{NC} ={1\over { {\tilde{h} }^3}} \int  e^{-\beta H^{NC}(\hat{q},\hat{p})} \, d^{3}\!\hat{q} \,  d^{3}\!\hat{p}~,
\end{equation}
which is written for a single particle, and which includes 
the  ${1\over { {\tilde{h} }^3}}$ factor, 
that was derived in Section 2. $H^{NC} (\hat{q},\hat{p})$ is the Hamiltonian of a 
NC classical system, expressed in terms of noncanonical 
coordinates and momenta.
According to the relations (18) and (19), and considering
      $\theta_{ij}={1\over2} {\epsilon_{ijk}\theta_k}$ and
      $\bar{\theta}_{ij}={1\over2}{\epsilon_{ijk}\bar{\theta}_{k}}$,
it is easy to write the integration measure in (20), up to the second order
in the NC parameters, as
\begin{equation}
 d^{3}\!\hat{q} \,  d^{3}\!\hat{p} = (1-{\theta\bar{\theta}  \over  8}) ~  d^{3}\!{q} \, d^{3}\!{p}~,
\end{equation}   
having assumed 
$\overrightarrow{\theta}=(0,0,\theta)$ and $\overrightarrow{\bar\theta}=(0,0,\bar\theta)$. 
If one writes $H^{NC} (\hat{q},\hat{p})$ in terms of
canonical coordinates and momenta (which can be done 
by applying (18) and (19) to the Hamiltonian), then the 
 appropriate
 expression for the partition function turns out to be 
\begin{eqnarray}
Q_1^{NC} & = & { {1\over  {h^3(1+ {\theta\bar{\theta}\over8 })}}
\int  e^{-\beta H^{NC}({q},{p})}
~~  (1-{\theta\bar{\theta}  \over  8}) 
  \, d^{3}\!{q} \,  d^{3}\!{p}~} \nonumber \\
& = & { 1\over  {h^3}  }   \int  e^{-\beta H^{NC}({q},{p})}
   ~~  (1-{\theta\bar{\theta}  \over  4})   \, d^{3}\!{q} \,  d^{3}\!{p}~.
\end{eqnarray}
where we used (13) and (21).
 $H^{NC} ({q},{p})$ is the Hamiltonian of a 
NC classical system, expressed in terms of canonical 
coordinates and momenta.
Thus, to find the partition function of a single particle
 of a NC classical system, it is enough to rewrite
 the Hamiltonian of the system in terms of
canonical classical observables, and plug it into the relation (22).
In the case when the basic constituents of 
the system are non-interacting, one can write the
classical partition function for a system of $N$ particles in a
3-dimensional NC phase space as
\begin{equation}
 Q_N^{NC}  = { 1\over  {N!}  }  [Q_1^{NC}]^N ,
\end{equation}
where  $1/N!$  is the Gibbs correction factor.

\section{\large{Classical Ideal Gas on NC Phase Space}}
Let us consider a system of $N$ identical molecules, assumed to be monoatomic
 (so that there are no internal degrees of motion to be considered), confined to a space
 of volume   $ V (=L^3)$   and in equilibrium at a 
temperature $T$. Since there are no intermolecular
 interactions to be taken into account, the Hamiltonian
 of a single molecule of the system,
in NC classical phase space, is simply given by 
\begin{equation}
                    H^{NC}(\hat{q}, \hat{p})   = {  \hat{p}_i  \hat{p}_i   \over{2m}  }~.
\end{equation}
Thus, from the relations (18) and (19), the Hamiltonian takes the following
form
\begin{equation}
                    H^{NC}(q,p)   =  {1\over {2m}}  \bigg(    p^2-  {1\over2} ~  \bar{\theta} L_z   +  {1\over 16} ~
                    \bar{\theta\,}^2   (x^2+y^2)   \bigg)~,
\end{equation}
where 
 $  \overrightarrow{\bar{\theta}}  \cdot   (  \overrightarrow{q}  \times  \overrightarrow{p}  ) =\bar{\theta} L_z $ and
 $  ( \overrightarrow{\bar{\theta}}  \times \overrightarrow{q}  )^2
=\bar{\theta\,}^2   (x^2+y^2)  $,
with the assumption
\begin{equation}
\overrightarrow{\theta}=(0,0,\theta) ~, ~~  
\theta_{ij}={1\over2} {\epsilon_{ijk}\theta_k}
\end{equation}
and 
\begin{equation}
\overrightarrow{\bar\theta}=(0,0,\bar\theta) ~, ~~
\bar{\theta}_{ij}={1\over2}{\epsilon_{ijk}\bar{\theta}_{k}}~.
\end{equation}
 Therefore, the partition
function reads 
\begin{equation}
                  Q_1^{NC}  =   { 1\over  {h^3}  }   \int  e^{-\beta  {1\over {2m}}  (    p^2-  {1\over2}  \bar{\theta} L_z   +  {1\over 16}   \bar{\theta\,}^2   (x^2+y^2)   )  }
                                         ~~  (1-{\theta\bar{\theta}  \over  4})   \, d^{3}\!{q} \,  d^{3}\!{p}~.
\end{equation}
By performing the integral, up to second order in the 
NC parameters, and using Eq. (23), one obtains
\begin{equation}
                  Q_N^{NC}  =   { V^N\over  {N! h^{3N}}  }   (2 \pi mKT) ^{3N\over 2}
                                         ~  (1-{\theta\bar{\theta}  \over  4})^N ~.
\end{equation}
As a consequence, the Helmholtz free energy is given by
\begin{equation}
                  A^{NC}  = -KT  \ln{(Q_N^{NC})}   =A+ NKT \theta\bar{\theta}/4~,
\end{equation}
where $A$ is the Helmholtz free energy in the ordinary (commutative) phase space.
 The complete thermodynamics of the ideal gas can be derived from (29) and (30)
 in a straightforward way. For instance,
\begin{eqnarray}
                  S^{NC}  &=& -\bigg(   {   {\partial {A^{NC}}   }  \over    {   \partial  T}   }   \bigg)_{_{N,V}}
                           =  S  -  NK \theta\bar{\theta}/4~,\\
                  \mu^{NC}  &=& \bigg(   {   {\partial {A^{NC}}   }  \over    {   \partial  N}   }   \bigg)_{_{V,T}}
                             =  \mu  +  KT \theta\bar{\theta}/4~,\\
                  P^{NC}  &=&  -   \bigg(   {   {\partial {A^{NC}}   }  \over    {   \partial   V  }   }   \bigg)_{_{N,T}}
                           = P ~,\\
                  U^{NC}  &=&  -   { \partial  \over  {  \partial { \beta }} }  \ln  {( Q_N^{NC}  )}   = U
                  ~,\\
                  C_V^{NC}  &=& \bigg(   {   {\partial {U^{NC}}   }  \over    {   \partial  T}   }   \bigg)_{_{N,V}}
                             = C_V ~,
\end{eqnarray}
where $S$, $\mu$, $P$, $U$ and $C_V$ are the usual thermodynamic
quantities, which
have been calculated in [14].

\section{\large{Extreme Relativistic Gas on NC Phase Space}}
Let us now consider an ideal extreme relativistic gas, 
consisting of $N$ monoatomic molecules with
energy-momentum relationship  $E=pc $ , $c$ being the speed of light, confined to a space of volume
 $ V (= L^3) $ and in equilibrium at temperature $T$. 
In the NC classical phase space, the Hamiltonian of a single
 molecule
 of the system is given by 
\begin{eqnarray}
H^{NC}(q,p)   &=&   c  \sqrt {  \hat {p}_i  \hat  {p}_i  } \nonumber   \\
         &=&   c  \sqrt  {    \bigg(   p_i+{1\over2}\bar\theta_{ij}q_j     \bigg) \bigg(       p_i+{1\over2}\bar\theta_{ik}q_k    \bigg)   }\nonumber \\
         &=&  pc  -  { {c \, \bar{\theta}L_z}\over {4p}  } + { {c \, \bar{\theta\,}^2}\over {32p}  }\,(x^2+y^2) -
                                                 {  {{c\,\bar{\theta\,}^2} L_z^2}\over {32p^3}   } + \it{O} {({\bar \theta\,}^3)},
\end{eqnarray}
where we used the conditions (26) and (27). In view of (22), 
the partition function can be readily obtained as
\begin{eqnarray}
Q^{NC}_{1} &=& \frac{1}{h^3} \int e^{-\beta \bigg( pc - \frac{c
{\bar{\theta}} L_z}{4 p} + \frac{c {\bar\theta}^2}{32 p} (x^2+y^2) -
\frac{c {\bar\theta}^2 {L_z}^2}{32 p^3} \bigg) }~ \bigg( 1 -
\frac{\theta
{\bar\theta}}{4} \bigg) d^3qd^3p\nonumber\\
&=& 8 \pi V \bigg( \frac{KT}{hc} \bigg)^3 \bigg( 1 - \frac{\theta
{\bar{\theta}}}{4} \bigg),
\end{eqnarray}
whence
\begin{equation}
                  Q_N^{NC}  =  {1\over{N!}}  (8\pi{V}) ^N  {  \bigg( { {KT} \over {hc} } \bigg)^{3N} }~ { \bigg(1-{\theta\bar{\theta}  \over  4}\bigg)^N  }          .
\end{equation}
Finally, the complete thermodynamics of the ideal extreme
relativistic gas, in NC classical phase space,
 would be similar to Eq. (30) to (35).
\section{\large{3-Dimensional Harmonic Oscillator on NC Phase Space}}
We shall now examine a system of $N$, practically independent,
3-dimensional harmonic oscillators. The Hamiltonian of each of them, 
up to second order in the NC parameters, is then given by
\begin{equation}
H^{NC}(\hat{q},\hat{p})={1\over{2m}}   (  \hat {p}_i  \hat  {p}_i   )
+  {1\over2}    {m{\omega}^2}   (  \hat {q}_i  \hat  {q}_i   ),
\end{equation}
or
\begin{eqnarray}
H^{NC}({q},{p})
&=&   {1\over{2m}}   (   p_i+{1\over2}\bar\theta_{ij}q_j     ) (
p_i+{1\over2}\bar\theta_{ik}q_k    )  +
{1\over2}    {m{\omega}^2}    (  q_i-{1\over2} \theta_{ij}p_j ) ( q_i-{1\over2} \theta_{ik}p_k    ) \nonumber\\
&=&  {1\over{2m}} \bigg( p^2 - { 1\over2 }\,{ \bar{\theta} L_z } +
{1\over16}\, { \bar{\theta\,}^2   (x^2+y^2)  }     \bigg)  \nonumber\\
& & +~ {1\over2} {m{\omega}^2}  \bigg(  q^2 - { 1\over2 }\,{ {\theta} L_z }
+ {1\over16}\, { {\theta}^2   (p_x^2+p_y^2)  }  \bigg),
\end{eqnarray}
where we used Eq. (26) and (27). Thus, by analogy with Eq. (22), we 
readily obtain the partition function for the single oscillator as
\begin{eqnarray}
Q_1^{NC}   &=& {1\over {h^3}}   \int  e^  { -{{\beta}\over{2m}} \big(
p^2 - { 1\over2 }\,{ \bar{\theta} L_z } + {1\over16}\, {
\bar{\theta\,}^2   (x^2+y^2)  }     \big)  - {{\beta}\over2}
{m{\omega}^2}  \big(  q^2 - { 1\over2 }\,{ {\theta} L_z } +
{1\over16}\, { {\theta}^2   (p_x^2+p_y^2)  }  \big)
}\nonumber\\
& & \times   (1-{\theta\bar{\theta}  \over  4})   \, d^{3}\!{q} \,  d^{3}\!{p}~  \\
&=&  {1\over { ( \beta\hbar\omega  )^3 }}   { 1\over {
(1-{\theta\bar{\theta}  \over  8})   }  }
                                        {(1-{\theta\bar{\theta}  \over  4}) }    \\
                                  &=&  {1\over { ( \beta\hbar\omega  )^3 }}   {(1-{\theta\bar{\theta}  \over  8})
                                  },
\end{eqnarray}
so that
\begin{equation}
                   Q_N^{NC}  = {1\over { ( \beta\hbar\omega  )^{3N} }}   {\bigg(1-{\theta\bar{\theta}  \over  8}\bigg)^N } ~.
\end{equation}
Therefore, the complete thermodynamic expressions for 
the NC 3-dimensional harmonic
oscillator will be given by:
\begin{equation}
                  A^{NC}  = -KT  \ln{(Q_N^{NC})}   =A+ NKT \theta\bar{\theta}/8~,
\end{equation}

\begin{equation}
                  S^{NC}  = -(   {   {\partial {A^{NC}}   }  \over    {   \partial  T}   }   )_{_{N,V}}   =  S  -  NK \theta\bar{\theta}/8~,
\end{equation}

\begin{equation}
                  \mu^{NC}  = (   {   {\partial {A^{NC}}   }  \over    {   \partial  N}   }   )_{_{V,T}}   =  \mu  +  KT \theta\bar{\theta}/8~,
\end{equation}

\begin{equation}
                  P^{NC}  =  -   (   {   {\partial {A^{NC}}   }  \over    {   \partial   V  }   }   )_{_{N,T}}   = P ~,
\end{equation}

\begin{equation}
                  U^{NC}  =  -   { \partial  \over  {  \partial { \beta }} }  \ln  {( Q_N^{NC}  )}   = U ~,
\end{equation}

\begin{equation}
                  C_V^{NC}  = (   {   {\partial {U^{NC}}   }  \over    {   \partial  T}   }   )_{_{N,V}}   = C_V ~.
\end{equation}

\section{\large{Conclusions}}
In this work, we studied the effects of a NC phase space on some
classical statistical systems. We presented a formulation 
to calculate 
the classical partition function according to the canonical ensemble
theory, up to the second order in the noncommutative
parameters, Eq. (22). In order to make the volume of NC phase space 
dimensionless, we used the uncertainly relations ((10) - (12)) and, 
then, found the expressions 
(13) and (15) for the three- and two-dimensional NC phase space, respectively. 
In Sections 4, 5 and 6, we considered the generalization of three 
well-known classical systems to a NC phase space.
  The results show that the terms arising from
the noncommutativity of phase space emerge in thermodynamic quantities
that are not measurable ($A$, $S$ and $\mu$). We conclude that,
 in classic statistical 
mechanics, the present experimental instruments cannot currently
detect the effects of noncommutativity, at least up to the second
order in the NC parameters. The final point that must be emphasized
 is about the entropy. It is known that the NC phase 
space is generally ``more ordered''
 than the corresponding commutative phase space. The 
reason is that, for a physical problem defined in a NC 
phase space, the degeneracy of the
 system decreases [15] (a related facet of this issue is 
the existence of ``exotic'' phases, such as 
striped phases, in NC models, which have been observed in 
numerical simulations 
of noncommutative scalar field theory: see Refs. [16, 17] and 
references therein). Therefore, one expects the entropy
 of the NC system to be reduced with respect to the commutative phase 
space. This intuitive expectation is indeed confirmed by relations (31) 
and (46).

\end{document}